\RequirePackage{lineno}
\documentclass[aps,prl,twocolumn,tightenlines,superscriptaddress,showpacs,byrevtex]{revtex4}
\usepackage{subfigure}
\usepackage{graphicx} 
\usepackage{dcolumn}  

\graphicspath{{ps}}

\begin{document}


%
\title{ \quad\\[1.0cm] Observation of an excited $\Omega^-$ baryon}

\noaffiliation
\affiliation{University of the Basque Country UPV/EHU, 48080 Bilbao}
\affiliation{Beihang University, Beijing 100191}
\affiliation{Brookhaven National Laboratory, Upton, New York 11973}
\affiliation{Budker Institute of Nuclear Physics SB RAS, Novosibirsk 630090}
\affiliation{Faculty of Mathematics and Physics, Charles University, 121 16 Prague}
\affiliation{Chonnam National University, Kwangju 660-701}
\affiliation{University of Cincinnati, Cincinnati, Ohio 45221}
\affiliation{Deutsches Elektronen--Synchrotron, 22607 Hamburg}
\affiliation{Duke University, Durham, North Carolina 27708}
\affiliation{University of Florida, Gainesville, Florida 32611}
\affiliation{Key Laboratory of Nuclear Physics and Ion-beam Application (MOE) and Institute of Modern Physics, Fudan University, Shanghai 200443}
\affiliation{Justus-Liebig-Universit\"at Gie\ss{}en, 35392 Gie\ss{}en}
\affiliation{Gifu University, Gifu 501-1193}
\affiliation{II. Physikalisches Institut, Georg-August-Universit\"at G\"ottingen, 37073 G\"ottingen}
\affiliation{SOKENDAI (The Graduate University for Advanced Studies), Hayama 240-0193}
\affiliation{Gyeongsang National University, Chinju 660-701}
\affiliation{Hanyang University, Seoul 133-791}
\affiliation{University of Hawaii, Honolulu, Hawaii 96822}
\affiliation{High Energy Accelerator Research Organization (KEK), Tsukuba 305-0801}
\affiliation{J-PARC Branch, KEK Theory Center, High Energy Accelerator Research Organization (KEK), Tsukuba 305-0801}
\affiliation{IKERBASQUE, Basque Foundation for Science, 48013 Bilbao}
\affiliation{Indian Institute of Science Education and Research Mohali, SAS Nagar, 140306}
\affiliation{Indian Institute of Technology Bhubaneswar, Satya Nagar 751007}
\affiliation{Indian Institute of Technology Guwahati, Assam 781039}
\affiliation{Indian Institute of Technology Hyderabad, Telangana 502285}
\affiliation{Indian Institute of Technology Madras, Chennai 600036}
\affiliation{Indiana University, Bloomington, Indiana 47408}
\affiliation{Institute of High Energy Physics, Chinese Academy of Sciences, Beijing 100049}
\affiliation{Institute of High Energy Physics, Vienna 1050}
\affiliation{Institute for High Energy Physics, Protvino 142281}
\affiliation{INFN - Sezione di Napoli, 80126 Napoli}
\affiliation{INFN - Sezione di Torino, 10125 Torino}
\affiliation{Advanced Science Research Center, Japan Atomic Energy Agency, Naka 319-1195}
\affiliation{J. Stefan Institute, 1000 Ljubljana}
\affiliation{Institut f\"ur Experimentelle Kernphysik, Karlsruher Institut f\"ur Technologie, 76131 Karlsruhe}
\affiliation{King Abdulaziz City for Science and Technology, Riyadh 11442}
\affiliation{Department of Physics, Faculty of Science, King Abdulaziz University, Jeddah 21589}
\affiliation{Korea Institute of Science and Technology Information, Daejeon 305-806}
\affiliation{Korea University, Seoul 136-713}
\affiliation{Kyoto University, Kyoto 606-8502}
\affiliation{Kyungpook National University, Daegu 702-701}
\affiliation{LAL, Univ. Paris-Sud, CNRS/IN2P3, Universit\'{e} Paris-Saclay, Orsay}
\affiliation{\'Ecole Polytechnique F\'ed\'erale de Lausanne (EPFL), Lausanne 1015}
\affiliation{P.N. Lebedev Physical Institute of the Russian Academy of Sciences, Moscow 119991}
\affiliation{Faculty of Mathematics and Physics, University of Ljubljana, 1000 Ljubljana}
\affiliation{Ludwig Maximilians University, 80539 Munich}
\affiliation{Luther College, Decorah, Iowa 52101}
\affiliation{University of Maribor, 2000 Maribor}
\affiliation{Max-Planck-Institut f\"ur Physik, 80805 M\"unchen}
\affiliation{School of Physics, University of Melbourne, Victoria 3010}
\affiliation{University of Mississippi, University, Mississippi 38677}
\affiliation{University of Miyazaki, Miyazaki 889-2192}
\affiliation{Moscow Physical Engineering Institute, Moscow 115409}
\affiliation{Moscow Institute of Physics and Technology, Moscow Region 141700}
\affiliation{Graduate School of Science, Nagoya University, Nagoya 464-8602}
\affiliation{Universit\`{a} di Napoli Federico II, 80055 Napoli}
\affiliation{Nara Women's University, Nara 630-8506}
\affiliation{National Central University, Chung-li 32054}
\affiliation{National United University, Miao Li 36003}
\affiliation{Department of Physics, National Taiwan University, Taipei 10617}
\affiliation{H. Niewodniczanski Institute of Nuclear Physics, Krakow 31-342}
\affiliation{Nippon Dental University, Niigata 951-8580}
\affiliation{Niigata University, Niigata 950-2181}
\affiliation{Novosibirsk State University, Novosibirsk 630090}
\affiliation{Osaka City University, Osaka 558-8585}
\affiliation{Pacific Northwest National Laboratory, Richland, Washington 99352}
\affiliation{Panjab University, Chandigarh 160014}
\affiliation{Peking University, Beijing 100871}
\affiliation{University of Pittsburgh, Pittsburgh, Pennsylvania 15260}
\affiliation{Punjab Agricultural University, Ludhiana 141004}
\affiliation{Theoretical Research Division, Nishina Center, RIKEN, Saitama 351-0198}
\affiliation{University of Science and Technology of China, Hefei 230026}
\affiliation{Showa Pharmaceutical University, Tokyo 194-8543}
\affiliation{Soongsil University, Seoul 156-743}
\affiliation{University of South Carolina, Columbia, South Carolina 29208}
\affiliation{Stefan Meyer Institute for Subatomic Physics, Vienna 1090}
\affiliation{Sungkyunkwan University, Suwon 440-746}
\affiliation{School of Physics, University of Sydney, New South Wales 2006}
\affiliation{Department of Physics, Faculty of Science, University of Tabuk, Tabuk 71451}
\affiliation{Tata Institute of Fundamental Research, Mumbai 400005}
\affiliation{Excellence Cluster Universe, Technische Universit\"at M\"unchen, 85748 Garching}
\affiliation{Department of Physics, Technische Universit\"at M\"unchen, 85748 Garching}
\affiliation{Department of Physics, Tohoku University, Sendai 980-8578}
\affiliation{Earthquake Research Institute, University of Tokyo, Tokyo 113-0032}
\affiliation{Department of Physics, University of Tokyo, Tokyo 113-0033}
\affiliation{Tokyo Institute of Technology, Tokyo 152-8550}
\affiliation{Tokyo Metropolitan University, Tokyo 192-0397}
\affiliation{Virginia Polytechnic Institute and State University, Blacksburg, Virginia 24061}
\affiliation{Wayne State University, Detroit, Michigan 48202}
\affiliation{Yamagata University, Yamagata 990-8560}
\affiliation{Yonsei University, Seoul 120-749}
  \author{J.~Yelton}\affiliation{University of Florida, Gainesville, Florida 32611} 
  \author{I.~Adachi}\affiliation{High Energy Accelerator Research Organization (KEK), Tsukuba 305-0801}\affiliation{SOKENDAI (The Graduate University for Advanced Studies), Hayama 240-0193} 
  \author{J.~K.~Ahn}\affiliation{Korea University, Seoul 136-713} 
  \author{H.~Aihara}\affiliation{Department of Physics, University of Tokyo, Tokyo 113-0033} 
  \author{S.~Al~Said}\affiliation{Department of Physics, Faculty of Science, University of Tabuk, Tabuk 71451}\affiliation{Department of Physics, Faculty of Science, King Abdulaziz University, Jeddah 21589} 
  \author{D.~M.~Asner}\affiliation{Brookhaven National Laboratory, Upton, New York 11973} 
  \author{H.~Atmacan}\affiliation{University of South Carolina, Columbia, South Carolina 29208} 
  \author{T.~Aushev}\affiliation{Moscow Institute of Physics and Technology, Moscow Region 141700} 
  \author{R.~Ayad}\affiliation{Department of Physics, Faculty of Science, University of Tabuk, Tabuk 71451} 
  \author{V.~Babu}\affiliation{Tata Institute of Fundamental Research, Mumbai 400005} 
% ...
\collaboration{The Belle Collaboration}

\vspace{1.0in}
\begin{abstract}
Using data recorded with the Belle detector, 
we observe a new excited hyperon, an $\Omega^{*-}$ candidate decaying
into $\Xi^0K^-$ and $\Xi^-K^0_S$ with a mass of $2012.4\pm0.7\ {\rm (stat)\pm\ 0.6\ (\rm syst)}\ {\rm MeV}/c^2$
and a width of $\Gamma=6.4^{+2.5}_{-2.0}\ {\rm(stat)}\pm1.6\ {\rm(syst)}\ {\rm MeV}$. 
The $\Omega^{*-}$ is seen primarily in $\Upsilon(1S), \Upsilon(2S)$, and 
$\Upsilon(3S)$ decays.

\end{abstract}


\maketitle


{\renewcommand{\thefootnote}{\fnsymbol{footnote}}}
\setcounter{footnote}{0}

The $\Omega^-$ comprises three strange quarks. Its excited states have proved difficult to find;
the Particle Data Group (PDG)~\cite{PDG} lists only one of them, the 
$\Omega(2250)$, in its summary tables and it has a mass almost 600 ${\rm MeV}/c^2$ higher than 
that of the ground state. In addition, the particle listings detail two
other states for which the evidence of existence is considered to be ``only fair'', and they
are at even higher masses. 
The gap in the spectrum 
is surprising as there are negative-parity orbital excitations of many other
baryons approximately 300 ${\rm MeV}/c^2$ above their respective ground states. 
A particular feature of $\Omega^-$ baryons are their zero isospin which means that 
$\Omega^{*-}\to\Omega^-\pi^0$ decays are 
highly suppressed and this restricts the possible decays of excited states, 
with the largest expected decay mode for low-lying states being to $\Xi K$.
Such decays are analogous to the $\Omega_c^0\to\Xi_c^+K^-$ decays recently discovered by 
the LHCb Collaboration~\cite{LHCb} and confirmed soon after by Belle~\cite{BelleY}. 

In this Letter, we present the results of a search for $\Omega^{*-}$ using a data sample of $e^+e^-$
annihilations recorded by the Belle detector~\cite{Belle} operating at the KEKB asymmetric-energy
$e^+e^-$ collider~\cite{KEKB}. The analysis concentrates on data taken with the accelerator energy
tuned for the production of the $\Upsilon(1S)$, $\Upsilon(2S)$, and $\Upsilon(3S)$ resonances, with 
integrated luminosities of 5.7~fb$^{-1}$, 24.9~fb$^{-1}$, and 2.9~fb$^{-1}$, respectively. 
The decays of these narrow 
resonances proceed via gluons, and it has long been known that they contain an enhanced
baryon fraction compared with continuum $e^+e^-\to q\bar{q}$ events~\cite{CLEOB,ARGUS,CLEOC}.

We search for excited $\Omega^-$ decays into $\Xi^0K^-$ and $\Xi^-\bar{K}^0$~\cite{CC},
with subsequent decays into $\Xi^-\to\Lambda\pi^-$, $\Xi^0\to\Lambda\pi^0$, $\bar{K^0}\to\pi^+\pi^-$, $\Lambda\to p\pi^-$
and $\pi^0\to\gamma\gamma$. 
An excited $\Omega^-$ would be expected to decay strongly, and 
with almost equal branching fractions, into the above
two decay modes which would likely dominate the decays of any $\Omega^{*-}$ with a mass between
the $\Xi K$ and
$\Xi(1530)K$ thresholds. 

The Belle detector was a large solid-angle spectrometer comprising six sub-detectors: 
the silicon vertex detector (SVD), the 50-layer central
drift chamber (CDC), the aerogel cherenkov counter (ACC), the time-of-flight scintillation counter (TOF),
the electromagnetic calorimeter (ECL, divided into the barrel ECL in the central region, and the forward
and backward endcaps
at smaller angles with respect to the beam axis), and the
$K^0_L$ and muon detector. 
A superconducting solenoid produces a 1.5 T magnetic field throughout the first five of these sub-detectors.
The detector is described in more detail in Ref.~\cite{Belle}. Two inner detector configurations were used. 
The first comprised a 2.0 cm radius
 beampipe and a 3-layer SVD, and the second a 1.5 cm radius beampipe and a 4-layer SVD 
and a small-cell inner CDC.

Charged particles,
$\pi^{\pm}, K^{-}$, and $p$, are selected using the
information from the tracking (SVD, CDC) and charged-hadron identification (CDC, ACC, TOF) systems combined into a likelihood,
${\cal L}(h1:h2) = {\cal L}_{h1}/({\cal L}_{h1} + {\cal L}_{h2})$
where $h_1$ and $h_2$ are $p$, $K$, and $\pi$ as appropriate. 
Kaon candidates are defined as those with ${\cal L}(K:\pi)>0.9$ and ${\cal L}(K:p)>0.9$, which is 
approximately $ 83\%$ efficient.
For protons the requirements are ${\cal L}(p:\pi)>0.2$ and ${\cal L}(p:K)>0.2$, while for charged pions
${\cal L}(\pi:p)>0.2$ and ${\cal L}(\pi:K)>0.2$; these requirements are approximately 99\% efficient.

The $\pi^0$ candidates are reconstructed from two neutral clusters detected in the ECL,
each consistent with being
due to a photon and having an energy greater than 30 ${\rm MeV}$ in the laboratory frame (for those in the 
endcap calorimeter, the energy threshold is increased to 50 ${\rm MeV})$.

Candidate $\Lambda\ (K_S^0)$ decays are
 made from
$p\pi^-\ (\pi^+\pi^-)$ pairs with a production vertex significantly
separated from the average interaction point (IP) and a reconstructed invariant mass within 3.5 (5.0) ${\rm MeV}/c^2$ of the peak values. 

Each $\Xi^-$ candidate is reconstructed by combining a $\Lambda$ candidates with a $\pi^-$ candidate. The vertex formed
from these two is required to be at 
least $0.35$~cm from the IP, to be a shorter distance from the IP than the $\Lambda$ decay vertex, and to signify a positive $\Xi^-$ flight distance.
The $\Xi^0\to \Lambda\pi^0$ reconstruction is complicated by the fact that the $\pi^0$ has
negligible vertex position information. 
Combinations of $\Lambda$ and $\pi^0$ candidates are made, and then assuming the IP to be production point of the 
$\Xi^0$, the sum of the $\Lambda$ and $\pi^0$ momenta is taken as the momentum vector of the 
$\Xi^0$ candidate. The intersection of this trajectory with the reconstructed $\Lambda$ trajectory is then found and
this position is taken as the decay location of the $\Xi^0$ hyperon. The $\pi^0$ is then re-made from the two photons,
using this location as its point of
origin.
The reconstructed invariant mass of the $\pi^0$ candidate must be within $10.8~{\rm MeV}/c^2$ of the nominal mass 
(approximately $94\%$ efficient). 
To reduce the large combinatorial background
the momentum of the $\pi^0$ candidate is required to be greater than 200 ${\rm MeV}/c$. 
Combinations are retained if they have a decay location of the $\Xi^0$ indicating a positive $\Xi^0$ path length of greater than 
2~cm but less than the distance between the $\Lambda$ decay vertex and the IP. The refitting of the $\pi^0$ at the reconstructed
$\Xi^0$ decay vertex improves the $\Xi^0$ mass resolution by around $15\%$.

\begin{figure}[htb]  
                                                                                                                 
\includegraphics[width=3.2in]{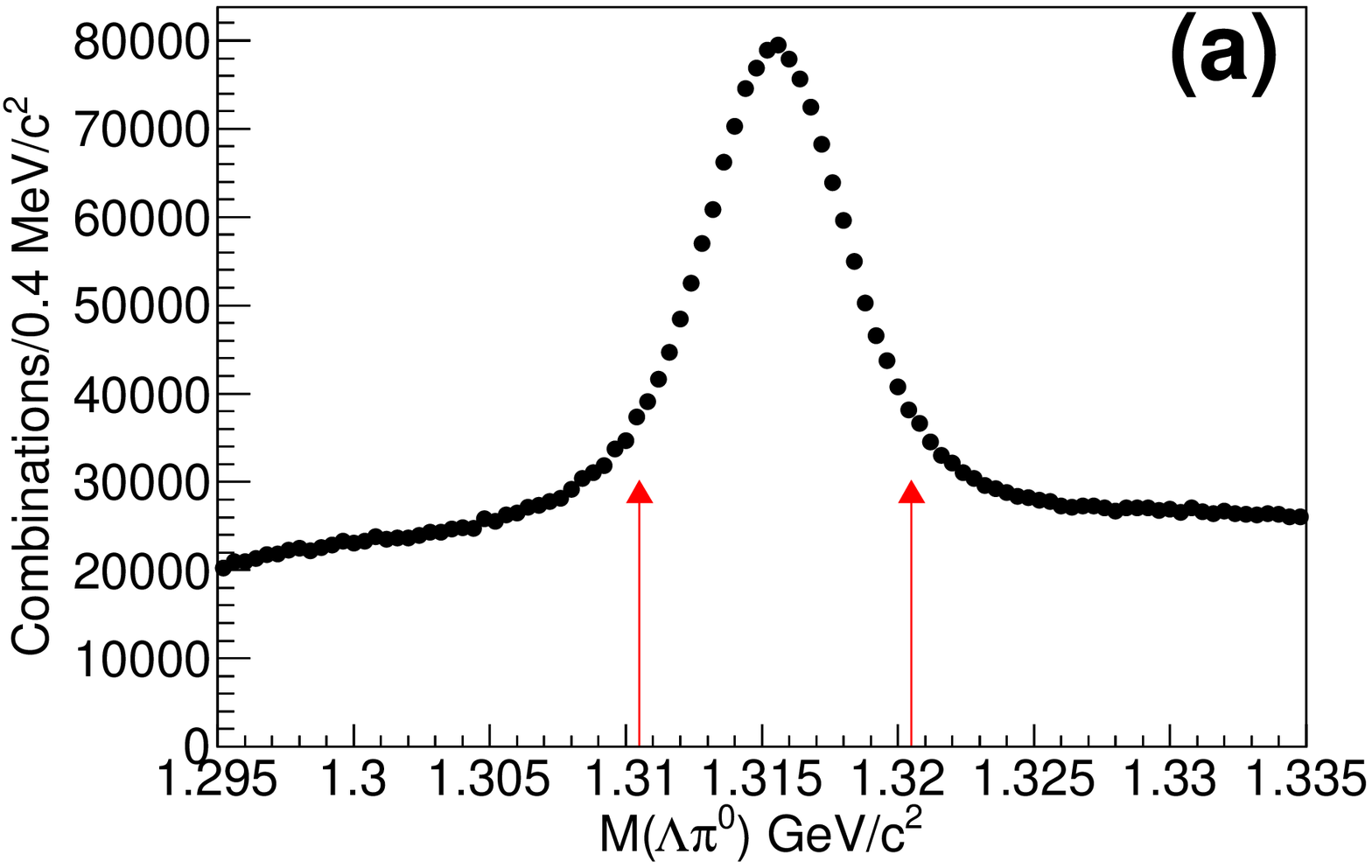}
\includegraphics[width=3.2in]{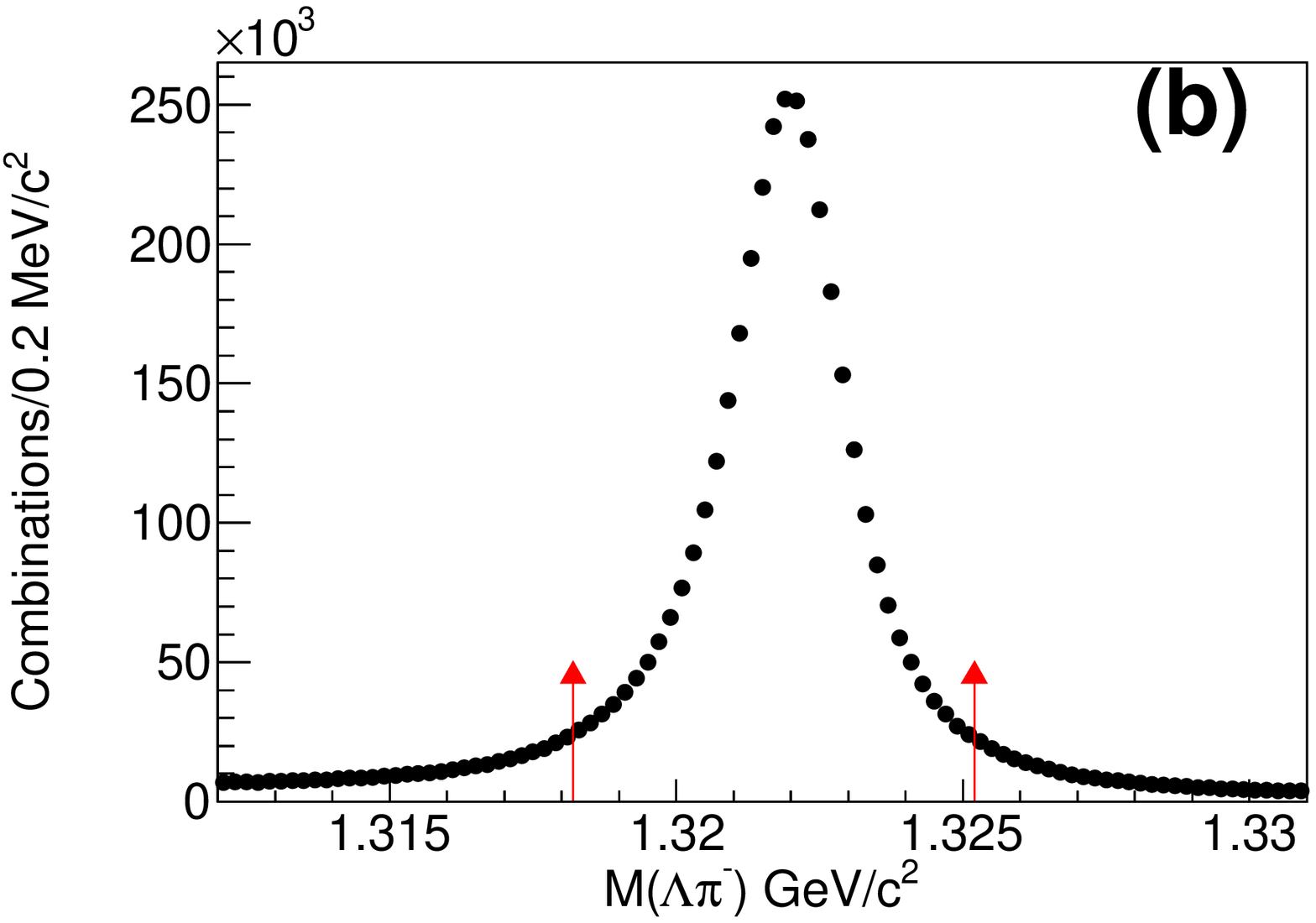}

\caption{Reconstructed invariant mass distributions of 
(a) $\Lambda\pi^0$ and (b) $\Lambda\pi^-$ combinations after all requirements. The arrows show the mass
windows used for $\Xi^0$ and $\Xi^-$ identification. }
\label{fig:fig1}
\end{figure}

The resultant invariant mass plots for the $\Xi^0$ and $\Xi^-$ candidates are shown in Fig.~\ref{fig:fig1}. 
The red vertical arrows indicate the limits of the reconstucted invariant masses of the candidates retained for further analysis,
which are $\pm 5.0~{\rm MeV}/c^2$ and $\pm 3.5~{\rm MeV}/c^2$ around the central values of the 
$\Xi^0$ and $\Xi^-$ mass peaks, respectively, which are each approximately 95\% efficient. 
For the $\Xi^0$ the value of the mass peak is $1.3155~{\rm GeV}/c^2$ 
and is higher than the PDG~\cite{PDG} value of $1.31486\pm0.00020~{\rm GeV}/c^2$. This difference is later used in the 
estimate of 
the systematic uncertainty of the $\Omega^{*-}$ resonance mass measurement.

The $\Xi^0$ and $\Xi^-$ candidates are kinematically constrained to their nominal masses~\cite{PDG}, 
and then combined with $K^-$ and $K^0_S$
candidates, respectively. The two particle combinations are kinematically constrained to come from a common vertex at the IP, and
the $\chi^2$ of this is required to be consistent with the daughters being produced by a common parent. For the $\Xi^0K^-$ case,
if there is more than one candidate with the same $\Lambda$ and $K^-$ but a different $\pi^0$, the one with the higher $\pi^0$ momentum is
kept and others discarded to avoid double counting. This occurs around $3\%$ of the time.

\begin{figure}[htb]  
                                                                                                                 
\hspace*{-.3in}\includegraphics[width=3.8in]{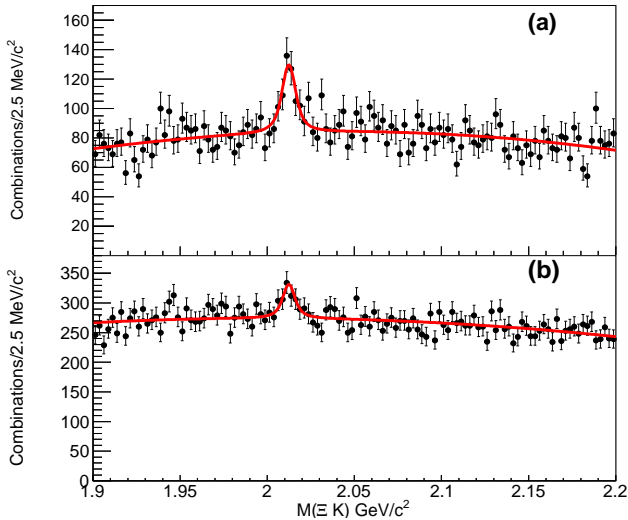}

\caption{
The (a) $\Xi^0 K^-$ and (b) $\Xi^-K^0_S$ invariant mass distributions in data 
taken at the $\Upsilon(1S), \Upsilon(2S)$, and $\Upsilon(3S)$ resonance energies. The curves show
a simultaneous fit to the two distributions with a common mass and width.  
}
\label{fig:fig2}
\end{figure}

Figure~\ref{fig:fig2} shows the $\Xi^0K^-$ and $\Xi^-K^0_S$ invariant mass distributions. Excesses are present in 
both distributions at around 2.01 ${\rm GeV}/c^2$. It should be noted that real $\Xi^0K^-$ combinations have three units of strangeness,
and are therefore highly suppressed. In contrast, $\Xi^-K^0_S$ combinations may have one unit of strangeness and thus have a larger
combinatorial  
background. 


A simultaneous fit applied to the two distributions is shown in Fig.~\ref{fig:fig2} and uses
fitting functions where the signal functions are 
Voigtian functions (Breit-Wigners convolved with a Gaussian resolution
functions) and the background functions second-order Chebyshev polynomials. The masses and intrinsic widths 
of the two Voigtian functions are kept the same.
The resolution functions are obtained from Monte Carlo (MC) events,
generated using EvtGen~\cite{EVTGEN} with the Belle detector response simulated using the GEANT3~\cite{GEANT3} framework,
and parameterized as Gaussian distributions with widths of $2.27~{\rm MeV}/c^2$ for $\Xi^0K^-$ and $1.77~{\rm MeV}/c^2$
for $\Xi^-K^0_S$.
The fit is made to the binned invariant mass distributions with a large number of small bins, using the
maximum-likelihood method.
A convenient test of the goodness-of-fit is the $\chi^2$ per degree of freedom ($\chi^2/{\rm d.o.f.}$)
for the distribution plotted in 2.5 ${\rm MeV}/c^2$ bins.
The signal yields, mass, instrinsic width, and $\chi^2/{\rm d.o.f.}$ resulting from this fit are listed in
Table~\ref{tab:Table1}.
We calculate the statistical significance of the signal by excluding the peaks from the fit,
finding the change in the log-likelihood ($\Delta[{\rm ln}(L)]$) and converting this to a p-value
taking into account the change in d.o.f. This is then converted to an
effective number of standard
deviations,  $n_{\sigma}$, and for this simultaneous fit we find $n_{\sigma}$ = 8.3.

Table~\ref{tab:Table1} also lists results obtained from fitting to each of the two distrubutions
separately. 
The signals in the $\Xi^0K^-$ and $\Xi^-K^0_S$ mass distributions have significances of $n_{\sigma} = 6.9$ and $n_{\sigma} = 4.4$,
respectively, and have statistically
compatible masses and widths.

\begin{table*}[htb]
\caption{The results of fits to the data shown in Fig.~\ref{fig:fig2}. The uncertainties shown are statistical only.}

\begin{tabular}
 { c | c | c | c| c |c |c }

\hline \hline
Data           &Mode & Mass (${\rm MeV}/c^2$)  & Yield & $\Gamma (\rm{MeV})$ & $\chi^2/{\rm d.o.f.}$ & $n_{\sigma}$  \\
\hline
$\Upsilon(1S,2S,3S)$ &   $\Xi^0K^-$, $\Xi^-K^0_S$  & $2012.4\pm0.7$  & $242\pm48$, $279\pm71$ & $6.4^{+2.5}_{-2.0}$ & 227/230  &   8.3    \\
                                       &  (simultaneous)&   &    &  &   &      \\
$\Upsilon(1S,2S,3S)$&  $\Xi^0K^-$ & $2012.6\pm0.8$ & $239\pm53$  & $6.1\pm2.6$ &    115/114      & 6.9          \\
$\Upsilon(1S,2S,3S)$&  $\Xi^-K^0_S$ &$2012.0\pm1.1$  &$286\pm87$   & $6.8\pm3.3$& 101/114       & 4.4 \\

\hline
Other&  $\Xi^0K^-$   & 2012.4 (Fixed) & $209\pm63$  & 6.4 (Fixed) &    102/116      & 3.4          \\
Other&  $\Xi^-K^0_S$ & 2012.4 (Fixed) & $153\pm89$  & 6.4 (Fixed) &    133/116      & 1.7          \\
\hline 
\hline
\end{tabular}

\label{tab:Table1}
\end{table*}

\begin{figure}[htb]  
                                                                                                                 
\hspace*{-.3in}\includegraphics[width=3.8in]{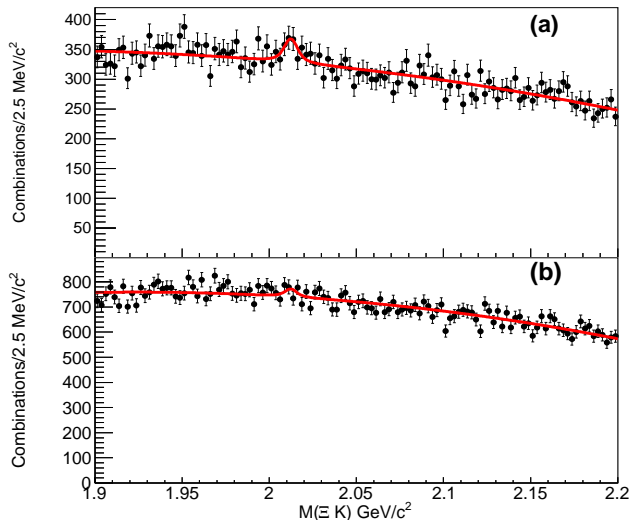}

\caption{ 
The (a) $\Xi^0 K^-$, (b) $\Xi^-K^0_S$ invariant mass distributions in data 
taken at energies other than $\Upsilon(1S), \Upsilon(2S)$, and $\Upsilon(3S)$ resonance energies. The curves show
the result of independent fits to the two distributions with masses and widths fixed to those found by the fit shown 
in Fig.~\ref{fig:fig2}. }

\label{fig:fig3}
\end{figure}

We have performed a series of checks to confirm the stability of the signal peak. Reasonable changes to the 
selection criteria of the daughter particles produce changes in the signal yield consistent with statistics. 
It would be surprising if an $\Omega^{*-}$ were not also produced in continuum $e^+e^-\to q{\bar q}$  events.
In Fig.~\ref{fig:fig3} we present mass distributions as in Fig.~\ref{fig:fig2} but for the remainder of the Belle 
data, which comprises a total of 946 ${\rm fb}^{-1}$ taken mostly at the $\Upsilon(4S)$ energy but also in the continuum
below and above this energy as well as at the $\Upsilon(5S)$. 
For the fits shown in Fig.~\ref{fig:fig3} we use second-order Chebyshev background functions
together with signal functions with mass and width fixed to the values found in the $\Upsilon(1S,2S,3S)$ data. Both 
distributions show excesses
in the signal region, and their statistical significances are listed in Table~\ref{tab:Table1}.

Taking into account the detection efficiency of the two modes, we use the results of the simultaneous fit to calculate the
branching fraction ratio 
${\cal R} = \frac{{\cal B}(\Omega^{*-}\to\Xi^0K^-)}{{\cal B}(\Omega^{*-}\to\Xi^-\bar{K}^0)} = 1.2 \pm 0.3$, where
statistical uncertainties dominate. Due to isospin symmetry this ratio would be expected to be 1, 
but the isospin mass-splitting of
the $\Xi$ and $K$ doublets will lead to an increase in this ratio of up to approximately 15\% 
depending on the spin associated with decay. Thus the obtained value of ${\cal R}$ is consistent with the 
expectation. 

The significance of the observation is largely unaffected by systematic uncertainties associated with the limited 
knowledge
of the resolution and momentum scale of the detector. However, the use of different background
functions can change the significance values.
If we replace the background functions by third-order Chebyshev polynomials, the significance of the 
signal in the simulataneous fit is  reduced to $n_{\sigma} = 7.2$. We take this value 
as the signal sigificance including systematic uncertainties.

The dominant systematic uncertainty of the mass measurement 
is that due to the masses of the $\Xi^0$ and $\Xi^-$ hyperons,
which enter almost directly into the calculation of the $\Omega^{*-}$ mass. Conservatively, we take the 
difference between the reconstructed $\Xi^0$ mass and the PDG value, 0.6 ${\rm MeV}/c^2$.
The Belle charged-particle momentum scale is very well understood, and the uncertainty in the $\Omega^{*-}$ mass measurement
due to this is
much smaller than 0.6 ${\rm MeV}/c^2$. 
Similarly, changing the fit function to a relativistic Breit-Wigner has negligible 
effect on the mass value.


MC simulation is known to reproduce the resolution of mass peaks within 10\% over a large number
of different systems. The resultant systematic uncertainty in $\Gamma$ from this source is ($\pm0.37\ {\rm MeV}$). 
Changing the
background shapes to third-order Chebyshev polynomials changes the measured value of $\Gamma$ by 1.6 ${ \rm MeV}$
and this is the dominant contributor to the systematic uncertainty of the width. 


The quark model~\cite{Capstick,Faustov,Loring,Roberts}, Skyrme model~\cite{Oh}, and lattice gauge theory~\cite{LGT} 
predict a
$J^P = \frac{1}{2}^-$ and $J^P = \frac{3}{2}^-$ pair of excited $\Omega^-$ states with masses in the 2000 ${\rm MeV}/c^2$
region. 
There are large discrepencies in the mass
predictions, but our value is in general closer to the those for the $J^P=\frac{3}{2}^-$ state. 
We also note that an $\Omega^{*-}$ with $J^P = \frac{3}{2}^-$ is restricted to decay to $\Xi K$ via
a d-wave, whereas a state with $J^P = \frac{1}{2}^-$ could decay via an s-wave. Thus the rather narrow width observed
implies that the $\frac{3}{2}^-$ identification is the more likely.

In summary, we have reported the observation of a new resonance, which we identify as an excited $\Omega^{-}$ baryon, 
found in the
decay modes $\Omega^{*-}\to\Xi^0K^-$ and $\Omega^{*-}\to\Xi^-K^0_S$. The measured mass of the resonance is 
[$2012.4\pm0.7\ {\rm (stat)}\pm0.6\ {\rm (syst)}]\ {\rm MeV}/c^2$ and its width, $\Gamma$, 
[$6.4^{+2.5}_{-2.0}{\rm (stat)}\pm1.6\ {\rm (syst)}]\ {\rm MeV}$].
It is found primarily in the decay of the narrow resonances $\Upsilon(1S)$, $\Upsilon(2S)$, and $\Upsilon(3S)$.

We thank the KEKB group for the excellent operation of the
accelerator; the KEK cryogenics group for the efficient
operation of the solenoid; and the KEK computer group,
the National Institute of Informatics, and the 
Pacific Northwest National Laboratory (PNNL) Environmental Molecular Sciences Laboratory (EMSL) computing group for valuable computing
and Science Information NETwork 5 (SINET5) network support.  We acknowledge support from
the Ministry of Education, Culture, Sports, Science, and
Technology (MEXT) of Japan, the Japan Society for the 
Promotion of Science (JSPS), and the Tau-Lepton Physics 
Research Center of Nagoya University; 
the Australian Research Council;
Austrian Science Fund under Grant No.~P 26794-N20;
the National Natural Science Foundation of China under Contracts
No.~11435013,  
No.~11475187,  
No.~11521505,  
No.~11575017,  
No.~11675166,  
No.~11705209;  
Key Research Program of Frontier Sciences, Chinese Academy of Sciences (CAS), Grant No.~QYZDJ-SSW-SLH011; 
the  CAS Center for Excellence in Particle Physics (CCEPP); 
Fudan University Grant No.~JIH5913023, No.~IDH5913011/003, 
No.~JIH5913024, No.~IDH5913011/002;                        
the Ministry of Education, Youth and Sports of the Czech
Republic under Contract No.~LTT17020;
the Carl Zeiss Foundation, the Deutsche Forschungsgemeinschaft, the
Excellence Cluster Universe, and the VolkswagenStiftung;
the Department of Science and Technology of India; 
the Istituto Nazionale di Fisica Nucleare of Italy; 
National Research Foundation (NRF) of Korea Grants No.~2014R1A2A2A01005286, No.2015R1A2A2A01003280,
No.~2015H1A2A1033649, No.~2016R1D1A1B01010135, No.~2016K1A3A7A09005 603, No.~2016R1D1A1B02012900; Radiation Science Research Institute, Foreign Large-size Research Facility Application Supporting project and the Global Science Experimental Data Hub Center of the Korea Institute of Science and Technology Information;
the Polish Ministry of Science and Higher Education and 
the National Science Center;
the Grant of the Russian Federation Government, Agreement No.~14.WW03.31.0026,
the Slovenian Research Agency;
Ikerbasque, Basque Foundation for Science, Basque Government (No.~IT956-16) and
Ministry of Economy and Competitiveness (MINECO) (Juan de la Cierva), Spain;
the Swiss National Science Foundation; 
the Ministry of Education and the Ministry of Science and Technology of Taiwan;
and the United States Department of Energy and the National Science Foundation.


\begin{thebibliography}{99}
\bibitem{CC}
Throughout this paper, 
the inclusion of the charge-conjugate decay mode is implied
unless otherwise stated.
\end{thebibliography}
\end{document}